\documentclass{acm_proc_article-sp}

\usepackage{graphicx}
\usepackage{subcaption}

\begin{document}

\title{Experiences with Automated Build and Test for Geodynamics Simulation Codes}

\numberofauthors{3}

\author{ 
        Eric M. Heien\textsuperscript{1}, Louise H. Kellogg\textsuperscript{1,2} \\
        \textsuperscript{1} Computational Infrastructure for Geodynamics\\
        \textsuperscript{2} Department of Geology\\
        University of California, Davis, USA
            \and
        Todd L. Miller\textsuperscript{3}, Becky Gietzel\textsuperscript{3}\\
        \textsuperscript{3} Center for High-Throughput Computing\\
        University of Wisconsin, Madison, USA
}

\maketitle

\begin{abstract}
The Computational Infrastructure for Geodynamics (CIG) is an NSF funded project that develops, supports, and disseminates community-accessible software for the geodynamics research community. CIG software supports a variety of computational geodynamic research from mantle and core dynamics, to crustal and earthquake dynamics, to magma migration and seismology.  To support this type of project a backend computational infrastructure is necessary.

Part of this backend infrastructure is an automated build and testing system to ensure codes and changes to them are compatible with multiple platforms and that the changes do not significantly affect the scientific results.  In this paper we describe the build and test infrastructure for CIG based on the BaTLab system, how it is organized, and how it assists in operations.  We demonstrate the use of this type of testing for a suite of geophysics codes, show why codes may compile on one platform but not on another, and demonstrate how minor changes may alter the computed results in unexpected ways that can influence the scientific interpretation.  Finally, we examine result comparison between platforms and show how the compiler or operating system may affect results.

\end{abstract}

\section{Introduction}
In geodynamics, computational experiments are often required to understand phenomena that occur over long time periods or at great depths.  Over the past several decades, the field of geodynamics has increasingly used computational simulations to model phenomena based on well established physics principles.  These simulations investigate issues such as how convection in the mantle affects the creation and subduction of tectonic plates \cite{JGRB:JGRB12209}, how tectonic deformation occurs over long time periods and causes earthquakes \cite{Williams:2005vz}, and how the composition of the earth affects global and regional scale seismic wave propagation \cite{Komatitsch:1999ki}, to name a few.

The research often involves a student or professor writing a computer code to solve a particular problem.  Even though these codes could potentially be used for related research, due to the cost of maintenance and support they are frequently abandoned.  This means that similar codes are often developed by multiple researchers, wasting time and money.  Furthermore, the time investment to write codes for parallel systems or use advanced computation technology (GPUs, vector processors, etc) means that most codes are relatively rudimentary and used only for specialized problems.

In response to this problem, the NSF funds the Computational Infrastructure for Geodynamics (CIG) to support a) reusable, well-documented geodynamics software that keeps pace with developments in computational technology; b) software building blocks for geodynamics from which state-of-the-art modeling codes can be effectively assembled; and c) strategic partnerships with the larger world of computational science and geoinformatics to ensure best practices in developing community-specific tool kits for scientific computation in solid-Earth sciences.

As part of these efforts, CIG has developed computational infrastructure to support each of these goals.  Recent work at CIG focused on creating a backend setup capable of automatically testing code and generating documentation to support users and developers.  There are multiple goals of the testing infrastructure including: a) automatic confirmation of successful compilation and execution on multiple platforms with different libraries; b) automatic validation of code output/results after changes to code; c) determining bounds of varying scientific results caused by differences in libraries, compilers, hardware, etc.  We believe the lessons learned from developing this infrastructure and working with domain scientists are applicable to other fields of science.

In Section \ref{sec-build-test} we outline the BaTLab system provided by the University of Wisconsin, Madison and how CIG interacts with it to automatically test supported codes.  In Section \ref{sec-vali-veri} we examine how code results may be affected by code changes or different platforms. Finally, in Section \ref{sec-conclusions} we examine what lessons may be applicable to other projects doing scientific code development.

\section{Automated Build and Test}
\label{sec-build-test}

\begin{figure}
\centering
\includegraphics[width=0.48\textwidth,trim=0in 3.8in 4.2in 0in,clip]{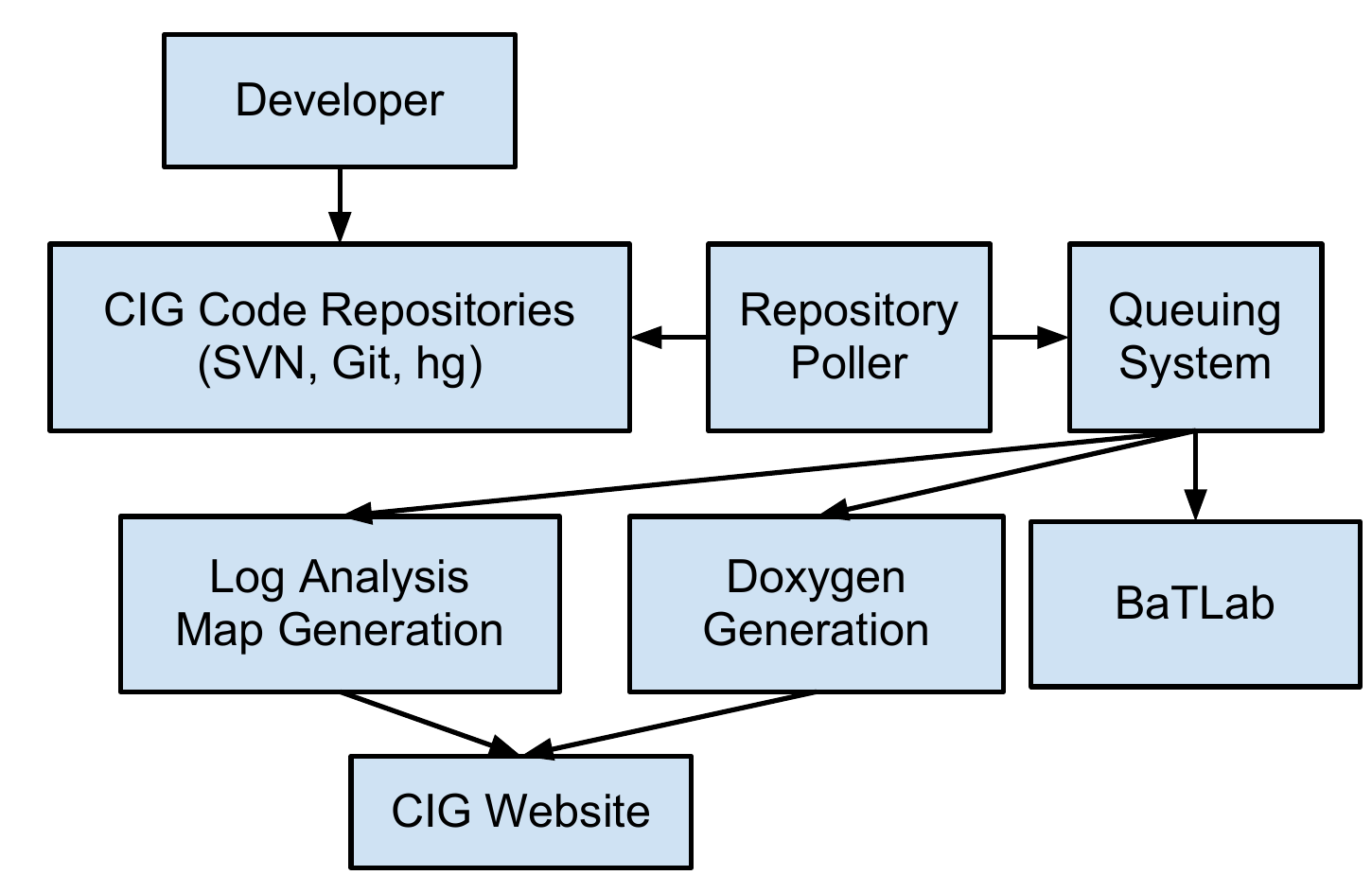}
\caption{Diagram of CIG backend organization.}
\label{fig-cig-backend}
\end{figure}

Figure \ref{fig-cig-backend} shows a general overview of the CIG backend infrastructure.  The key element is the developer who makes changes to a repository.  A backend polling mechanism periodically checks these repositories for changes and adds commands to a queueing system.  The commands include analyzing logs and generating maps of user locations, creating doxygen documentation for updated versions of code, and creating/submitting code tests to the BaTLab setup.  A polling/queuing setup rather than a trigger mechanism is used to avoid losing commands due to machine downtime on the repositories, backend, or testing system.  In this paper we only examine the BaTLab part of this diagram.

\begin{figure}
\centering
\includegraphics[width=0.48\textwidth,trim=0in 3.2in 4.6in 0in,clip]{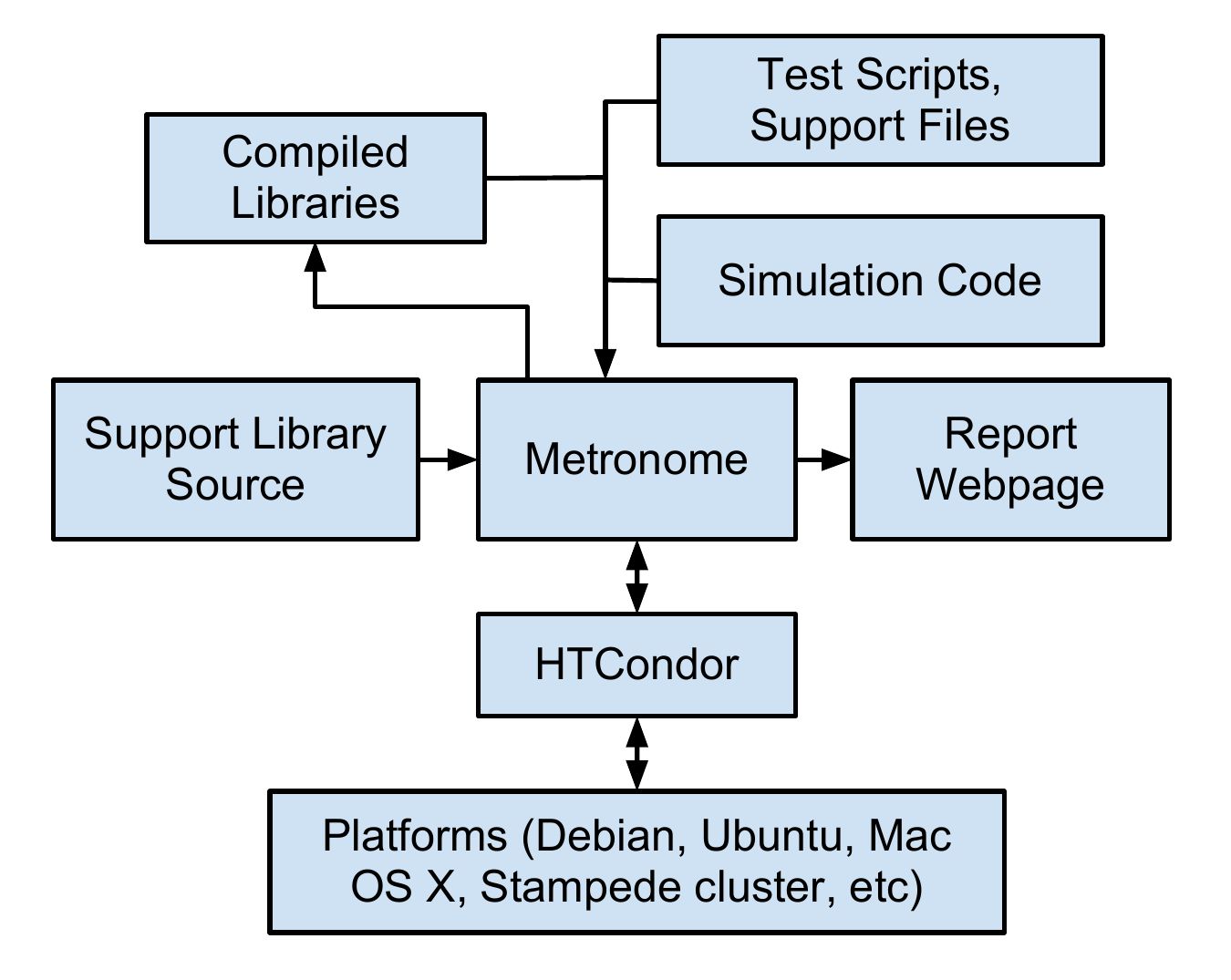}
\caption{Diagram of BaTLab.}
\label{fig-batlab-diagram}
\end{figure}

CIG uses the BaTLab (Build and Test Lab) system \cite{Pavlo:2006:NBT:1267793.1267814}, developed at the University of Wisconsin at Madison, to help perform automated build and testing.  The details of the BaTLab setup are shown in Figure \ref{fig-batlab-diagram}.  The main interface to the hardware platforms on BaTLab is the Metronome automated build and test system, which interacts with the hardware through the HTCondor distributed computing platform.  The BaTLab system used in this work consists of 15 platforms running on multiple hardware configurations.  The use of the HTCondor system allows sharing of resources among multiple users, sophisticated scheduling and resource allocation, and easy addition of external computing resources.  In particular, this allowed extension of the testing framework to the XSEDE TACC Stampede system where many CIG supported codes are used.

Most codes require support libraries that may not be installed on the operating system so it is necessary to provide precompiled libraries for each code.  To avoid recompiling libraries during each test run, libraries are precompiled through the BaTLab system and stored for use during actual test runs.  This also allows fast and easy testing of codes with different versions of libraries, and simple updating of libraries to new versions and new platforms.

When a CIG code is submitted to BaTLab for testing several steps occur to configure the system, run the tests, and report the results.  First, the necessary Metronome input files and scripts are dynamically generated for the code from a central database of CIG codes.  This database indicates where the code is stored, what libraries and support files are needed, what platforms to use, and what tests to run.  Dynamically generating test scripts from the database makes it easier to add and support many codes and libraries, avoid redundant files common to multiple codes, and isolate the testing logic in a single place.

\begin{figure*}[!t]
\centering
\includegraphics[width=0.98\textwidth]{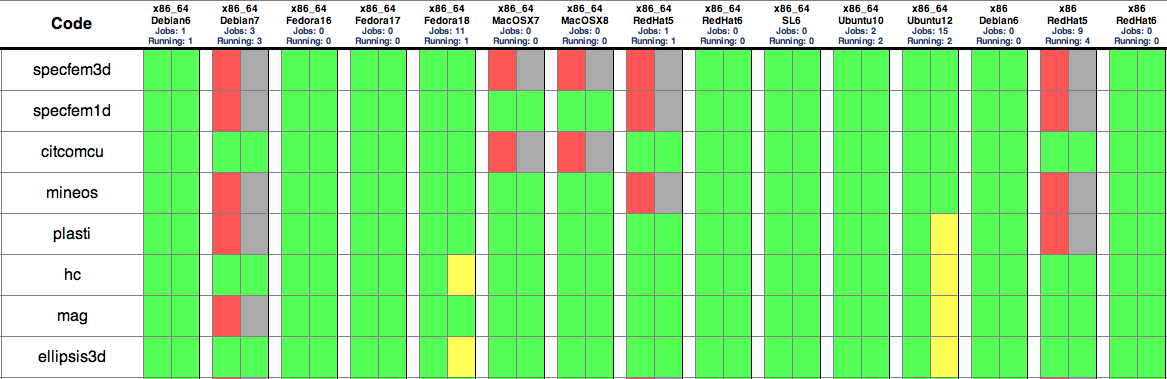}
\caption{BaTLab testing result dashboard. Red indicates a failed build or test, yellow indicates a build/test that is still running, green indicates a success. Grey indicates a test that couldn't run because of a failed build.}
\label{fig-testing-dashboard}
\end{figure*}

Code tests on BaTLab are split into two phases: the build phase - setting up necessary libraries and compiling the code - and the test phase - running the compiled code and checking the outputs.  If the build phase fails the test phase is not run. The results of compiling and testing a code are reported both through email and a web dashboard.  Figure \ref{fig-testing-dashboard} shows the dashboard after testing 8 codes on 15 platforms.  The rows show the code being tested and the columns indicate the platform.  The colors indicate the results of the build and test phases.

CIG has configured 15 geodynamics codes to run on the BaTLab system, out of 23 total supported codes.  The additional codes have not yet been added due to complex library and/or compilation requirements.  However, the 15 configured codes are a good representation of geodynamics simulations, ranging from seismic wave propagation (e.g. SPECFEM variants), mantle convection (Citcom variants, Ellipsis3D, HC), short-term crustal dynamics (RELAX, SELEN), long-term tectonics (Plasti) and magnetohydrodynamics simulation of the geodynamo (Mag, Calypso).  

\subsubsection*{Lessons Learned}
Based on these tests, CIG has discovered several issues with the supported codes that have been or remain to be corrected.  The most common issue is that many codes are written to work with a particular compiler and use features or flags that are not supported by other compilers.  This appears to be particularly true for Fortran based codes where there are significant differences between supported flags between compilers.  This problem is best solved by using an automated configuration script (such as provided through autoconf or CMake) to test the compiler for flag compatibility rather than using hardcoded flags or compiler names.

Another minor issue is the variability of compiler support for language constructs.  What is considered valid code for some compilers may produce warnings or errors with other compilers, especially for the GNU vs. Portland Group vs. Intel compilers.  Although this is most common for differences of supported language extensions (C++03 vs. C++11 or F77 vs. F95 vs. F03) it also applies to issues among all code such as unused variables, type qualifier issues or NULL references.  These are more difficult to solve through configuration scripts since the issues will depend on the context.  The best solution is to compile the code with each compiler and manually address whatever warnings or errors appear.

\section{Validation and Verification}
\label{sec-vali-veri}
A key aspect of scientific code development is validation and verification of the codes and their outputs. Many scientific codes are developed with minimal testing of their correctness, no evaluation of the differences in output on multiple platforms or compilers, and no way of tracking how subsequent changes to the code affect the results. In this area the meaning of verification will vary depending on the physics being simulated, the calculations performed, and the output generated.  Here we examine two case studies involving verifying the results of codes that represent some of the key simulations performed by CIG supported codes.

\subsection{Changes to Simulation}

\begin{figure}[!t]
\begin{subfigure}[h]{0.5\textwidth}
\centering
\includegraphics[width=0.48\textwidth]{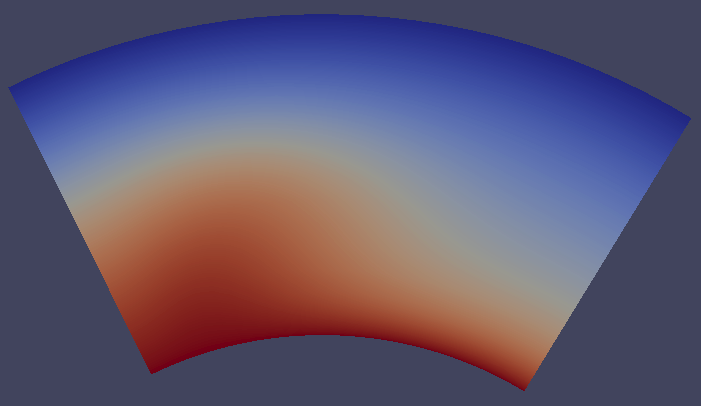} \includegraphics[width=0.49\textwidth]{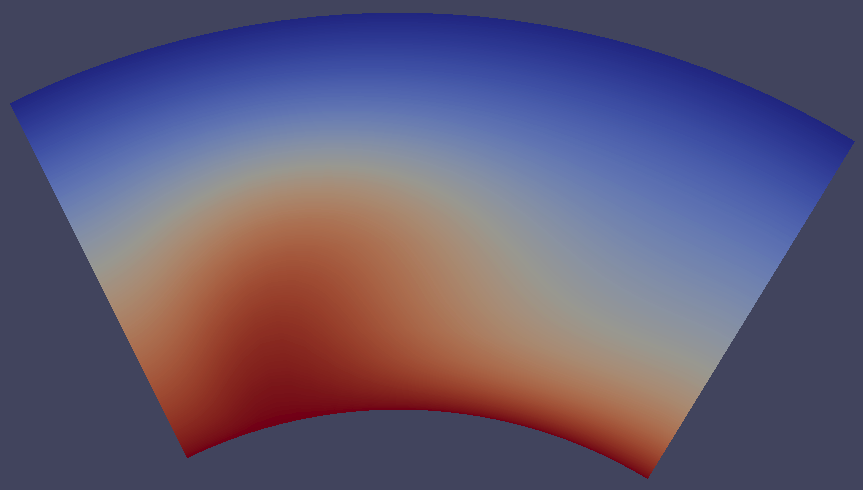}
\caption{T = 50.}
\label{fig-plume-comparison-a}
\end{subfigure}
\begin{subfigure}[b]{0.5\textwidth}
\centering
\includegraphics[width=0.48\textwidth]{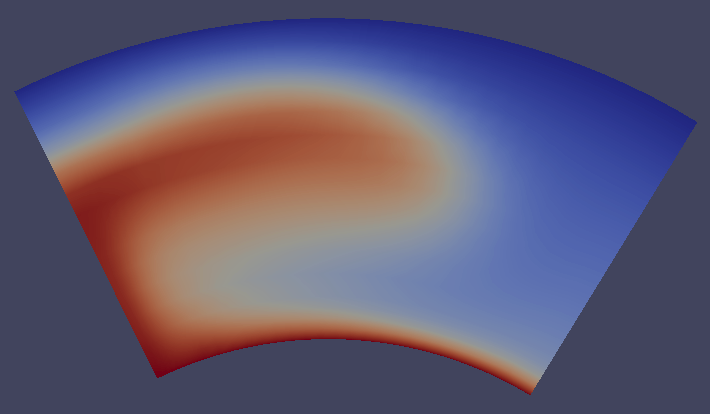} \includegraphics[width=0.485\textwidth]{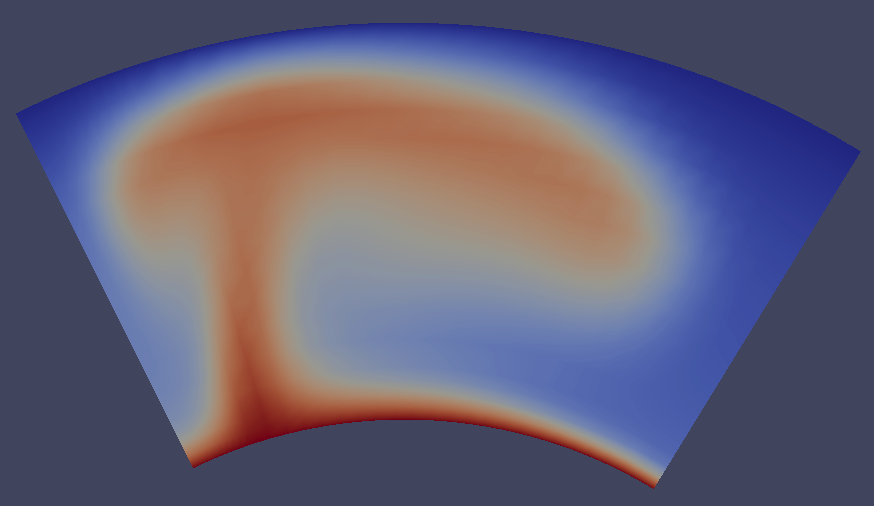}
\caption{T = 150.}
\label{fig-plume-comparison-b}
\end{subfigure}
\begin{subfigure}[b]{0.5\textwidth}
\centering
\includegraphics[width=0.49\textwidth]{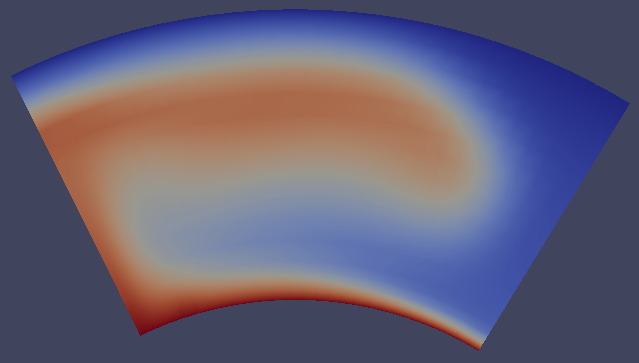} \includegraphics[width=0.472\textwidth]{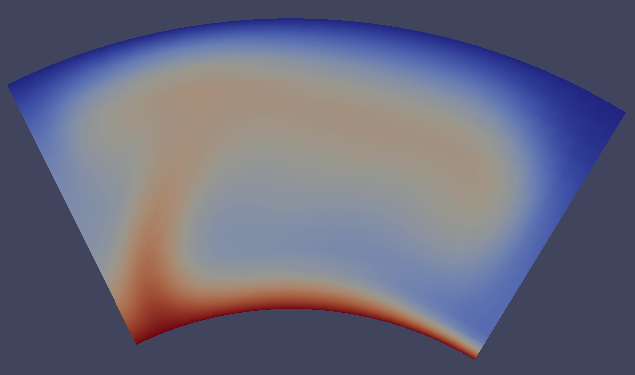}
\caption{T = 250.}
\label{fig-plume-comparison-c}
\end{subfigure}
\caption{An example of how a minor change to the finite element solver altered the simulation physics. The left figures show the original convection plume development, and the right figures show the new development.}
\label{fig-plume-comparison}
\end{figure}

CitcomS \cite{JGRB:JGRB12209,GGGE:GGGE831} is a finite element code written in C that solves for thermo-chemical convection within a spherical shell.  It uses a finite element approximation to the relevant equations and solves the approximation with an iterative sparse solver.  The velocity and temperature fields are propagated forward in time such that a change in initial conditions or the approximation may cause unexpected effects later.

During development of CitcomS a change was made to the iterative solver to make it more accurate by continuing solver iterations until not only the temperature and pressure solutions converged but also the velocity continuity. However, this change resulted in unexpected effects on evolution of simple plume advection as shown in Figure \ref{fig-plume-comparison}. In this Figure the original results are on the left and new results on the right.  Near the beginning of the simulation at $T=50$ (Figure \ref{fig-plume-comparison-a}), the developing plumes are visually similar in shape.  By $T=150$ (Figure \ref{fig-plume-comparison-b}) there is clear divergence with one plume separating from the side of the domain.  At $T=250$ (Figure \ref{fig-plume-comparison-c}) the results are clearly distinct from each other and it is unclear which simulation result is valid.

\paragraph*{Lessons Learned}
The changes that caused this were on the order of 3-4 lines of code, and the developers were confident that the tighter convergence criterion would not significantly affect the results.  At this time CitcomS did not have an automated result verification system and it was only by chance that this difference was discovered.  Even basic result verification would likely not have detected it since commonly used comparison metrics (boundary heat flux, momentum equation force) were not significantly different between the two simulations.

This clearly indicates the need for rigorous automated testing to be performed on all developed scientific codes.  Even if a change to the code does not affect results for a given setup, due to the complex nature of most scientific codes there is no guarantee that other setups will not be affected. Such tests should be composed while the code is developed rather than afterwards, and any bug that is discovered should have a test composed to ensure it does not reappear later.

\subsection{Effects of Different Platforms}
\begin{figure}
\centering
\includegraphics[width=0.47\textwidth]{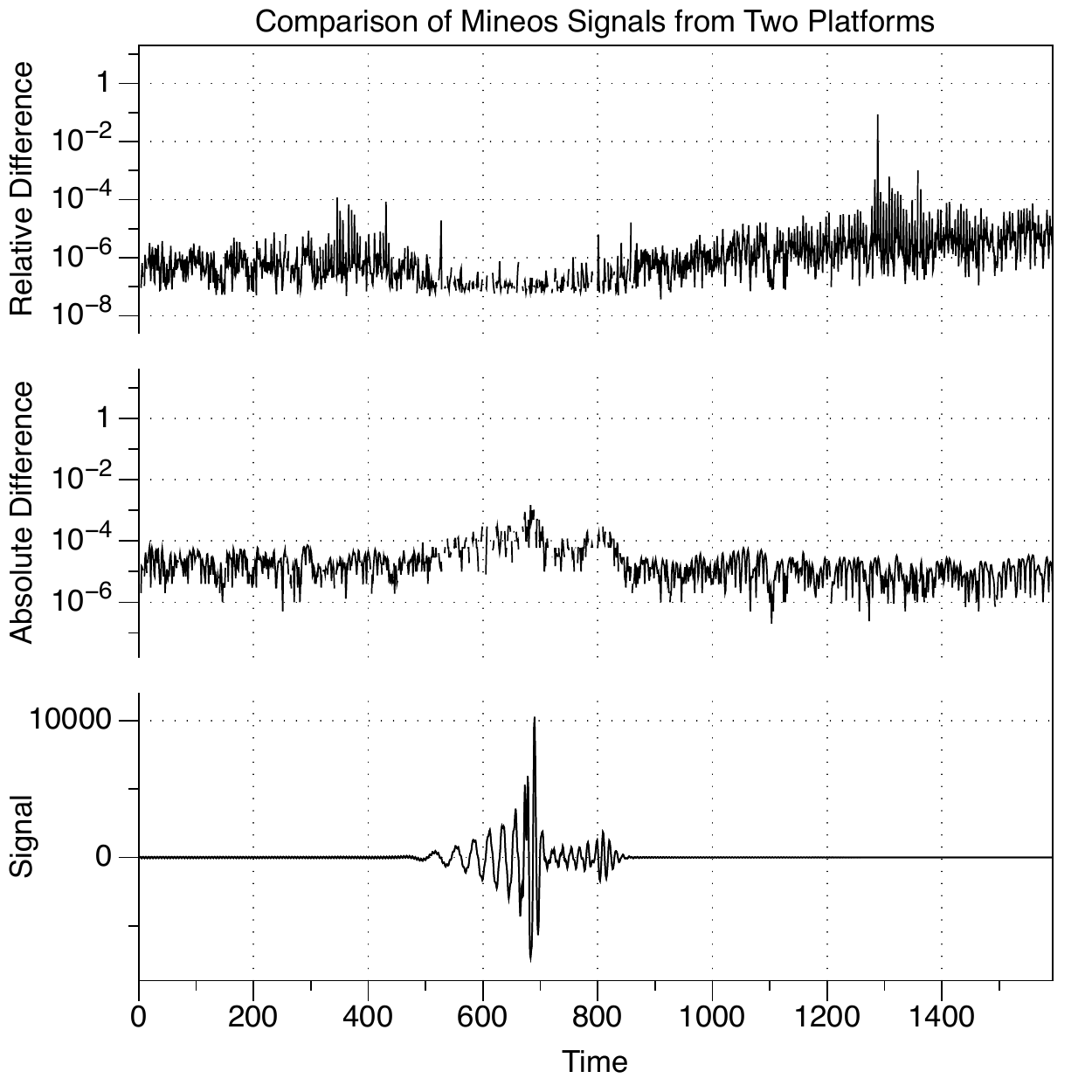}
\caption{Comparison of Mineos calculation on (A) a laptop with an Intel Core i7 compiled using gfortran, and (B) a cluster with Intel Xeon E5 compiled using ifort.}
\label{fig-mineos-multi-platform}
\end{figure}

Most of the codes supported by CIG have some discrepancy between results calculated on different platforms.  This is caused by minor variations in hardware or software implementation of floating point calculations between platforms.  Here we look at one of the larger differences found in the Mineos code.  Mineos computes synthetic seismograms at specified locations in a spherically symmetric non-rotating Earth by summing normal modes.  Results will vary depending on the ordering of summation, hardware evaluation of the functions comprising the normal modes, and other data storage or manipulation issues introduced by compilers.

Figure \ref{fig-mineos-multi-platform} shows two example signals from Mineos computed on two different platforms.  The bottom graph shows the original signal, the middle graph shows the absolute difference and the top graph shows the relative difference.  Even though both simulations were compiled and run using the exact same configuration and runtime options, the results are significantly different.  The greatest relative difference occurs after the main event (the earthquake wave) has passed, so that the signal of interest has tapered off to almost nothing. That is important because the difference does not, in this particular example, affect the scientific interpretation of the model. The automated testing reveals the difference, and the domain scientist determines whether this difference is significant for the problem under consideration.

\paragraph*{Lessons Learned}
Appropriate methods to compare results for scientific codes are highly application dependent. Since there are minor variations between platforms an exact diff of the output files is not possible. Visually comparing the results is generally not possible because of the automated nature of the system.  Because there are several orders of magnitude difference between the signal at a weak point vs. a strong point, an absolute value comparison will need to have a high cutoff making it less valuable for comparison.  Also, because of the varying signal at each measurement point, dozens of cutoffs would need to be determined to handle each output data set.  As seen in Figure \ref{fig-mineos-multi-platform}, relative difference also does not work well because weaker points in the signal can have very high relative differences (up to 9\% in this case).

In the case of seismogram generating codes we have found a normalized coefficient of correlation to be a good measure of output similarity.  This will automatically adapt to any sort of signal and will find a correlation of nearly 1 for similar results from the tested codes.  For finite element codes other comparisons such as $l^2$-norm may be more appropriate, and CIG is working to develop tools to assist with this. 

\section{Conclusions}
\label{sec-conclusions}
Ensuring the validity of scientific codes and developing automated verification systems is an arduous but very important task. This step can be difficult for domain researchers, likely because they do not have the tools or expertise to make such comparisons. We have demonstrated how an automated verification systems can be used to develop a testing infrastructure for use by the scientific community. We use this system to examine why codes may not compile on all platforms, and how minor changes in code can affect the results.

\bibliographystyle{abbrv}
\bibliography{wssspe_paper}

\end{document}